\pdfoutput=1

\documentclass{article}
\usepackage{amsmath, graphicx}

\usepackage[a4paper]{geometry}
\usepackage{authblk}
\usepackage{longtable}

\usepackage[sort]{cite}

\title{Predictive modelling of training loads and injury in Australian football}
\author[1]{David L. Carey\thanks{d.carey@latrobe.edu.au}}
\author[2]{Kok-Leong Ong}
\author[3]{Rod Whiteley}
\author[1]{Kay M. Crossley}
\author[1,4]{Justin Crow}
\author[1,5]{Meg E. Morris}
\affil[1]{La Trobe Sport and Exercise Medicine Research Centre, College of Science, Health and Engineering, La Trobe University, Melbourne, Australia}
\affil[2]{SAS Analytics Innovation Lab, La Trobe Business School, La Trobe University, Melbourne, Australia}
\affil[3]{Aspetar Orthopedic and Sports Medicine Hospital, Doha, Qatar}
\affil[4]{Essendon Football Club, Melbourne, Australia}
\affil[5]{Healthscope, Northpark Private Hospital, Melbourne, Australia}

\date{}

\begin{document}

\maketitle

\begin{abstract}
\noindent
To investigate whether training load monitoring data could be used to predict injuries in elite Australian football players, data were collected from elite athletes over 3 seasons at an Australian football club. Loads were quantified using GPS devices, accelerometers and player perceived exertion ratings. Absolute and relative training load metrics were calculated for each player each day (rolling average, exponentially weighted moving average, acute:chronic workload ratio, monotony and strain). Injury prediction models (regularised logistic regression, generalised estimating equations, random forests and support vector machines) were built for non-contact, non-contact time-loss and hamstring specific injuries using the first two seasons of data. Injury predictions were generated for the third season and evaluated using the area under the receiver operator characteristic (AUC). Predictive performance was only marginally better than chance for models of non-contact and non-contact time-loss injuries (AUC$<$0.65). The best performing model was a multivariate logistic regression for hamstring injuries (best AUC=0.76). Learning curves suggested logistic regression was underfitting the load-injury relationship and that using a more complex model or increasing the amount of model building data may lead to future improvements. Injury prediction models built using training load data from a single club showed poor ability to predict injuries when tested on previously unseen data, suggesting they are limited as a daily decision tool for practitioners. Focusing the modelling approach on specific injury types and increasing the amount of training data may lead to the development of improved predictive models for injury prevention.

\end{abstract}

\newpage
\setcounter{page}{1}
\section{Introduction}

Training loads are associated with injury in team sport athletes \cite{rogalski2013,colby2014,hulin2015,carey2016,malone2016,murray2016,thornton2016}. Monitoring and adjusting loads to reduce injury risk is considered an important aspect of athlete management \cite{soligard2016}, especially since injuries can have a detrimental effect on team sport performance \cite{hagglund2013}. Measuring relative and absolute training loads has been recommended for athlete monitoring \cite{drew2016,soligard2016}. Relative training loads have been quantified using the acute:chronic workload ratio \cite{blanch2016,gabbett2016,hulin2015,murray2016,williams2016better}, and absolute training loads are commonly reported using cumulative totals or absolute weekly loads \cite{rogalski2013,colby2014}. Other methods of training load calculation that take into account daily variations, such as monotony and strain, have been proposed as useful for athlete monitoring \cite{foster2001}. Bittencourt et al. suggested that modelling complex interactions between different risk factors should be considered when building injury prediction models, and that machine learning approaches may be appropriate \cite{bittencourt2016}.

The ability of training load monitoring to predict future injury is not well established. Few studies have examined the performance of training load models to predict injuries in new data \cite{gabbett2010,thornton2016}. No model has established superiority for accurate injury predictions, and the ability of models to predict particular injury types is unknown. Techniques to deal with the large imbalance between injured and non-injured samples, and the volume of data needed to build accurate predictive models have not been explored.

We aimed to investigate the ability of training loads to predict future injuries in elite Australian football players. Relative and absolute training loads, player ages and session types were used as predictors. Univariate and multivariate predictive models were trained on two years of player monitoring data and evaluated on one year of unseen future data. Models were compared using the area under the receiver operator characteristic (AUC). Different data processing protocols to deal with collinear predictors and unbalanced classes were compared, and learning curves were constructed to explore how the amount of available data influenced the quality of future predictions.

\section{Methods}

\subsection{Participants}

The participants involved in the study were from one professional Australian football club. The club fielded 45, 45 and 43 players in the 2014, 2015 and 2016 seasons respectively, giving 133 player seasons from 75 unique athletes. Informed consent was received from the club for collection and analysis of de-identified training and injury data. The project was approved by the La Trobe University Faculty of Health Sciences Human Ethics Committee (FHEC14/233).

In the 2016 season the club participating in this study fielded an uncommonly high number of new players due to multiple season long suspensions. The impact of this on the predictive models was explored by comparing the results for new versus returning players. This enabled evaluation of the impact of introducing new players on the performance of injury models.

\subsection{Data collection}

Player tracking data were collected using commercially available 10 Hz global positioning system (GPS) devices and 100 Hz triaxial accelerometers (Catapult Optimeye S5). All players were monitored during all field-based training sessions and matches. The devices used in the study have been validated for use in this athletic population \cite{boyd2011,rampinini2015}. Additionally, players gave a rating of perceived exertion (RPE) after each session \cite{foster2001}. Missing data were imputed using predictive mean matching \cite{buuren2011}.

Club medical staff recorded all injuries. Injuries were classified using the Orchard Sports Injury Classification System (OSICS) \cite{rae2007} and were categorised as contact or non-contact. Injury severity was classified as either transient (not causing unavailability for training or matches) or time-loss (causing the player to be unavailable for regular training or match activity). Hamstring injuries were defined to include all injures when the OSICS `specific' category was `Hamstring strain' or `Hamstring tendon injury'.

\subsection{Training load quantification}

Training loads were quantified using five different training load variables (Table 1). For each workload variable we constructed rolling averages and exponentially weighted moving averages over 3, 6 and 21 days, as well as monotony and strain over 7 days \cite{foster2001}. Acute:chronic workload ratios with 3 and 6 day acute time windows and 21 day chronic time windows were derived \cite{carey2016}, as well as exponentially weighted workload ratios \cite{murray2016,williams2016better} (see Appendix I for detailed calculations). Time frames for load calculations were based on a previous study of the relationship between training loads and injury in a similar cohort \cite{carey2016}.

\begin{table}[h]
	\centering
	\caption{Workload variables and descriptions.}
	\label{tab1}
	\begin{tabular}{cc}
		\textbf{Training load variable}                    & \textbf{Definition}                                                                                                    \\ \hline
		Distance (m)                                       & Distance above 3 km/h                                                                                                  \\
		Moderate speed running (MSR) (m)                   & Distance between 18-24 km/h                                                                                            \\
		High speed running (HSR) (m)                       & Distance above 24 km/h                                                                                                 \\
		Session-RPE (arbitrary units) \cite{foster2001} & \begin{tabular}[c]{@{}c@{}}Rating of perceived exertion multiplied by\\ session duration\end{tabular}                  \\
		Player load (arbitrary units) \cite{boyd2011}   & \begin{tabular}[c]{@{}c@{}}Proprietary metric measuring the magnitude of\\ rate of change of acceleration\end{tabular} \\ \hline
	\end{tabular}
\end{table}

\subsection{Modelling approach}

Predictive models were built on two years of load and injury data (model training data) and tested on one season of unseen future data (testing data) (Figure 1). Evaluating models on a season of unseen data is required to get an estimate of the ability to predict injuries. Multivariate statistical models can have many degrees of freedom and can be tuned to fit a particular data set very well. To test whether a model will generalise and be useful in practice, the predictions must be evaluated on a new data set \cite{kuhn2013}.

\begin{figure}[h]
	\centering
	\includegraphics[width = 0.8\textwidth]{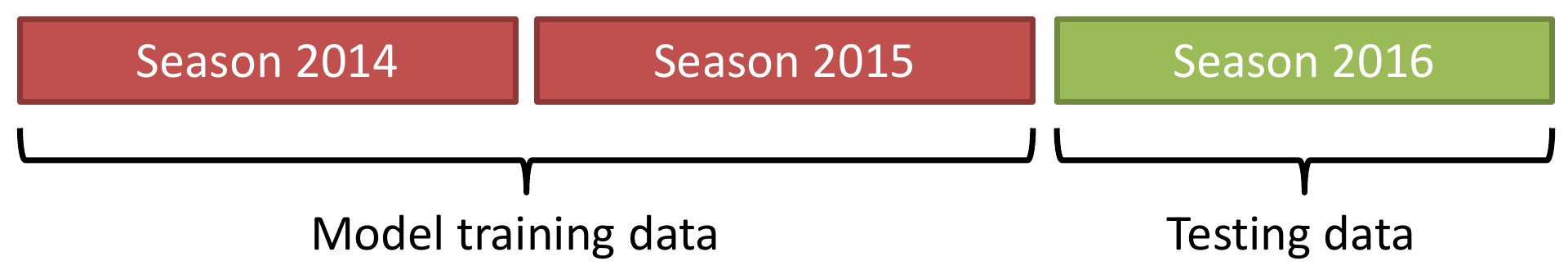}
	\caption{Data split for model training and testing.}
	\label{fig1}
\end{figure}

Models were built to try and predict whether an athlete would be injured, given knowledge of their training loads. Each day that a player completed training or a match, or reported an injury, was included as an observation in the model training data. Predictions were then generated for each training or match day in the testing data set and evaluated against actual injury incidence.

Three different types of injury were considered; any non-contact (NC), non-contact causing time-loss (NCTL) and any hamstring injury (HS). Separate models were built for each injury outcome to investigate whether predictive models performed better on specific injury types. Hamstring injuries were chosen as they were the most frequently occurring specific injury in this cohort and are a common injury in Australian football \cite{orchard2013}. A possible lag period between spikes in training load and increased injury risk has been reported in previous studies \cite{hulin2015}. Models were also built for the likelihood of a player sustaining an injury in the next five days to investigate whether including a lag period could improve predictive performance.

\subsection{Prediction algorithms}

Multiple algorithms were tested to compare their ability to predict injury in team sport athletes. The multivariate models (constructed with all training load variables) were:
\begin{itemize}
	\item Logistic regression (LR) is commonly used to model injury outcomes \cite{colby2014,gabbett2010,murray2016}. Regularisation was introduced due to the large number of predictor variables used \cite{zou2005}.
	\item Random forest (RF) models were chosen for their ability to fit non-linear patterns in data and deal with collinear predictors \cite{breiman2001}. Random forests have been used in injury prediction studies in rugby league \cite{thornton2016}.
	\item Generalised estimating equations (GEE) are an extension of generalised linear models that account for correlations between repeated observations taken from the same participant \cite{liang1986}. A binomial link function and auto-regressive correlation structure was used \cite{williamson1996}.
	\item Support vector machines (SVM) with a radial basis function were used to model the potentially non-linear pattern between load and injury in high dimensional data \cite{vapnik1998}.
\end{itemize}

In addition, univariate LR models were constructed for each training load variable \cite{colby2014,gabbett2010,malone2016} to provide a comparison for the more complex multivariate and non-linear models used.

All analyses were performed using the R statistical programming language \cite{rcore}. Model parameters were tuned using 10-fold cross validation on the model training data.

\subsection{Data pre-processing}

To allow for players to accumulate sufficient training loads to calculate workload ratios and exponentially weighted moving averages, the first 14 days of each season were removed from the model training and testing data. This loss of data could potentially be avoided in future studies if monitoring of chronic workloads extended into the off-season period. Players in rehabilitation training were excluded from modelling until they returned to full training.

Many of the training load variables collected were likely to be correlated, thus our prediction problem may suffer from multi-collinearity. Principal component analysis (PCA) is a dimensionality reduction process that reduces a large number of predictor variables to a smaller number of uncorrelated variables \cite{kuhn2013} and has been employed in studies of training load monitoring \cite{williams2016}. Each multivariate model was trained with unprocessed data and data pre-processed with PCA. The number of components was determined using a 95\% variance cut-off threshold \cite{kuhn2013}.

Class imbalance refers to prediction problems when one class is far more common than the other \cite{kuhn2013}. Injury prediction suffers from large class imbalance since injuries are far less common than days when a player doesn't get injured. Severe class imbalance can cause prediction algorithms to have trouble correctly predicting the rare class \cite{kuhn2013}. Two sampling techniques were implemented to combat class imbalance. Under-sampling randomly removes non-injury days from the model building data until there is an equal number of injury and non-injury days. Synthetic minority over-sampling (SMOTE) synthetically creates new injury samples in the model training data and under-samples a fraction of the non-injury days to even up the classes \cite{chawla2002}. Each model was built using unprocessed data, under-sampled data and SMOTE sampled data.

\subsection{Model evaluation}

To evaluate the performance of each modelling approach, predicted injury probabilities were generated for each training or match day in the testing data. A receiver operator characteristic curve was constructed and the AUC calculated. A perfect model would have AUC=1.0 and random guessing would be expected to produce AUC=0.5. This gives a performance metric preferable to error rate for problems with unbalanced data; where low error can be achieved by simply predicting the more common class every time \cite{kuhn2013} (e.g. if the injury rate is 1\%, a model always predicting no-injury is 99\% accurate).

To account for uncertainties introduced into the modelling procedure by randomly sampling the data during the pre-processing and model tuning stages, 50 simulations of the entire process were run, generating a set of performance estimates for each model.

\section{Results}

The number of reported injuries (and frequency relative to the number of sessions) is shown in Table~2. Hamstring injury rates were similar across seasons (0.004 vs. 0.003 injuries per session), however non-contact (0.035 vs. 0.014 injuries per session) and non-contact time-loss (0.017 vs. 0.009 injuries per session) injury rates were lower. Descriptive statistics for each of the training load variables are contained in Supplementary Table I. Training load variables were generally similar across the training and testing data sets.

\begin{table}[h]
	\centering
	\caption{Injury counts and rates (relative to total number of sessions) in the model training and testing data.}
	\label{tab2}
	\begin{tabular}{ccc}
		\textbf{Injury outcome}           & \textbf{Model training data (2014 \& 2015)} & \textbf{Testing data (2016)} \\ \hline
		NC                                & 321 (0.035)                                 & 67 (0.014)                   \\
		NCTL                              & 156 (0.017)                                 & 42 (0.009)                   \\
		HS                                & 36 (0.004)                                  & 13 (0.003)                   \\
		NC-lag                            & 1601 (0.174)                                & 479 (0.103)                  \\
		NCTL-lag                          & 784 (0.085)                                 & 295 (0.063)                  \\
		HS-lag                            & 183 (0.020)                                 & 88 (0.019)                   \\
		Total records (match \& training) & 9203                                        & 4664                         \\ \hline
	\end{tabular}
\end{table}

\subsection{Predictive ability of different modelling approaches}

Predictive performance was limited for multivariate models when using un-processed data (Figure 2). Using regularised LR to model hamstring injuries performed best (mean AUC=0.72), all other multivariate models had a mean AUC of less than 0.65 on the testing data. Univariate models performed worse than multivariate models for each injury outcome (best AUC$<$0.6 for NC and NCTL, and best AUC$<$0.7 for HS).

Attempting to account for a possible lag time between training load spikes and injury occurrence (NC lag, NCTL lag and HS lag in Figure 2) did not improve predictive performance (mean AUC=0.50-0.57). In general, models provided predictions only marginally better than chance.

The performance of HS injury models (particularly LR and RF) was more variable than non-specific injury models (Figure 2). The increased variability is likely due to the smaller number of HS injuries in the model training data (n=36). The random inclusion or exclusion of one or more of these injuries during the model building stage has a greater impact on the final model.

No particular prediction algorithm showed a strong tendency to outperform others across different injury outcomes. The more complex models (RF and SVM) did not tend to outperform generalised linear models (LR). Accounting for individual clustering effects (GEE) did not lead to better results.

\begin{figure}[h]
	\centering
	\includegraphics[width = 0.6\textwidth]{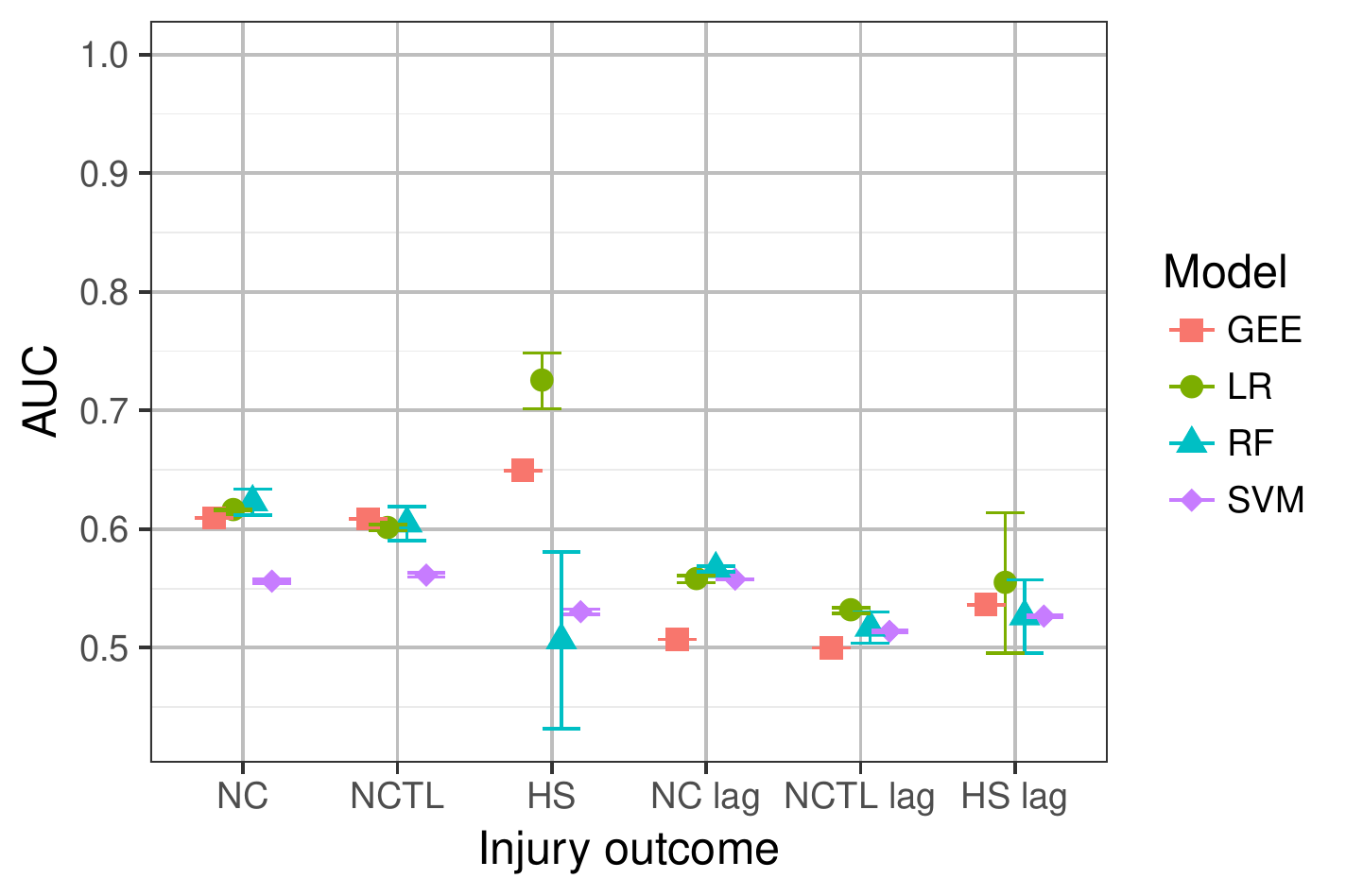}
	\caption{Area under ROC curve evaluated on the testing data set (mean and standard deviation of 50 simulations) for different prediction algorithms and injury outcomes (no data pre-processing).}
	\label{fig2}
\end{figure}

\subsubsection{Hamstring injury prediction}

Regularised LR for hamstring injuries showed better performance (mean AUC=0.72) than other models (Figure 2). The model gives a predicted injury probability each day, how this translates into practise (i.e. modifying player training and preventing injuries) is dependent on the preferences of the decision maker. Specifically, how they weight the consequences (cost) of a false negative (missed injury) relative to a false positive prediction (unnecessarily cancelling or modifying a session by decreasing volume or intensity). To illustrate this we considered three arbitrary relative costs: 1 missed injury (i.e. one we didn't predict) costs as much as 50 unnecessarily modified sessions (aggressive risk), as much as 100 sessions (moderate risk), or as much as 1000 sessions (conservative risk). The estimated optimal performance metrics for the three choices are shown in Table 3. There is a trade-off between correctly predicting more injuries and incorrectly flagging non-injury sessions. In general, the AUC values for other models in the study were similar or below the HS model, suggesting they would not perform significantly better than the results in Table 3.

\begin{table}[h]
	\centering
	\caption{Estimated optimal performance of HS injury models for different relative cost ratios (values reported as median of 50 simulations).}
	\label{tab3}
	\begin{tabular}{ccccccc}
		\textbf{\begin{tabular}[c]{@{}c@{}}Cost of false\\ negative\\ relative\\ to false\\ positive\end{tabular}} & \textbf{\begin{tabular}[c]{@{}c@{}}True\\ positive\\ rate\end{tabular}} & \textbf{\begin{tabular}[c]{@{}c@{}}False\\ positive\\ rate\end{tabular}} & \textbf{\begin{tabular}[c]{@{}c@{}}Positive\\ likelihood\\ ratio\end{tabular}} & \textbf{\begin{tabular}[c]{@{}c@{}}Negative\\ likelihood\\ ratio\end{tabular}} & \textbf{\begin{tabular}[c]{@{}c@{}}Probability\\ injury\\ given\\ positive\\ prediction\end{tabular}} & \textbf{\begin{tabular}[c]{@{}c@{}}Probability\\ injury\\ given\\ negative\\ prediction\end{tabular}} \\ \hline
		\begin{tabular}[c]{@{}c@{}}50\\ (aggressive\\ risk)\end{tabular}                                           & 0.08                                                                    & 0.004                                                                    & 17.9                                                                           & 0.93                                                                           & 0.05                                                                                                  & 0.003                                                                                                 \\
		\begin{tabular}[c]{@{}c@{}}100\\ (moderate\\ risk)\end{tabular}                                            & 0.54                                                                    & 0.11                                                                     & 5.0                                                                            & 0.52                                                                           & 0.01                                                                                                  & 0.001                                                                                                 \\
		\begin{tabular}[c]{@{}c@{}}1000\\ (conservative\\ risk)\end{tabular}                                       & 0.92                                                                    & 0.53                                                                     & 1.7                                                                            & 0.16                                                                           & 0.005                                                                                                 & 0.0004                                                                                                \\ \hline
	\end{tabular}
\end{table}

\subsubsection{Effect of data pre-processing}

The effects of different data pre-processing protocols are shown in Figure 3. Model performance varied under different protocols, yet the differences in predictive ability were generally small. Reducing the number of predictors to a smaller, uncorrelated set by applying PCA pre-processing caused minor performance improvements in the models considered (Figure 3). This suggested that multicollinearity was a potential cause of poor performance when using un-processed data. Additionally, the variability in performance tended to decrease (especially for RF models).

Under-sampling non-injury days led to performance decreases for all models except the SVM (Figure~3). This is possibly due to the information lost from the model training data when a large number of the non-injury days are removed. Under-sampling may not be appropriate for the injury prediction problem. SMOTE sampling did not lead to any major performance improvements (Figure 3). In the SMOTE procedure new injury observations were synthetically created using the common characteristics observed in actual injuries. This may not help the generalisability of models if new injuries show little resemblance to past ones. SVM models were the exception, and benefited from both sampling methods used, suggesting their performance was more negatively affected by imbalance between injury and non-injury observations. Combining SMOTE sampling and PCA pre-processing was similarly unsuccessful in improving predictive performance.

Applying sampling methods to the data to try and reduce the amount of imbalance between the number of injured and non-injured observations led to increased variability in results (Figure 3). This is a consequence of randomly removing different subsets of the data before building models in each simulation.

\begin{figure}[h]
	\centering
	\includegraphics[width = \textwidth]{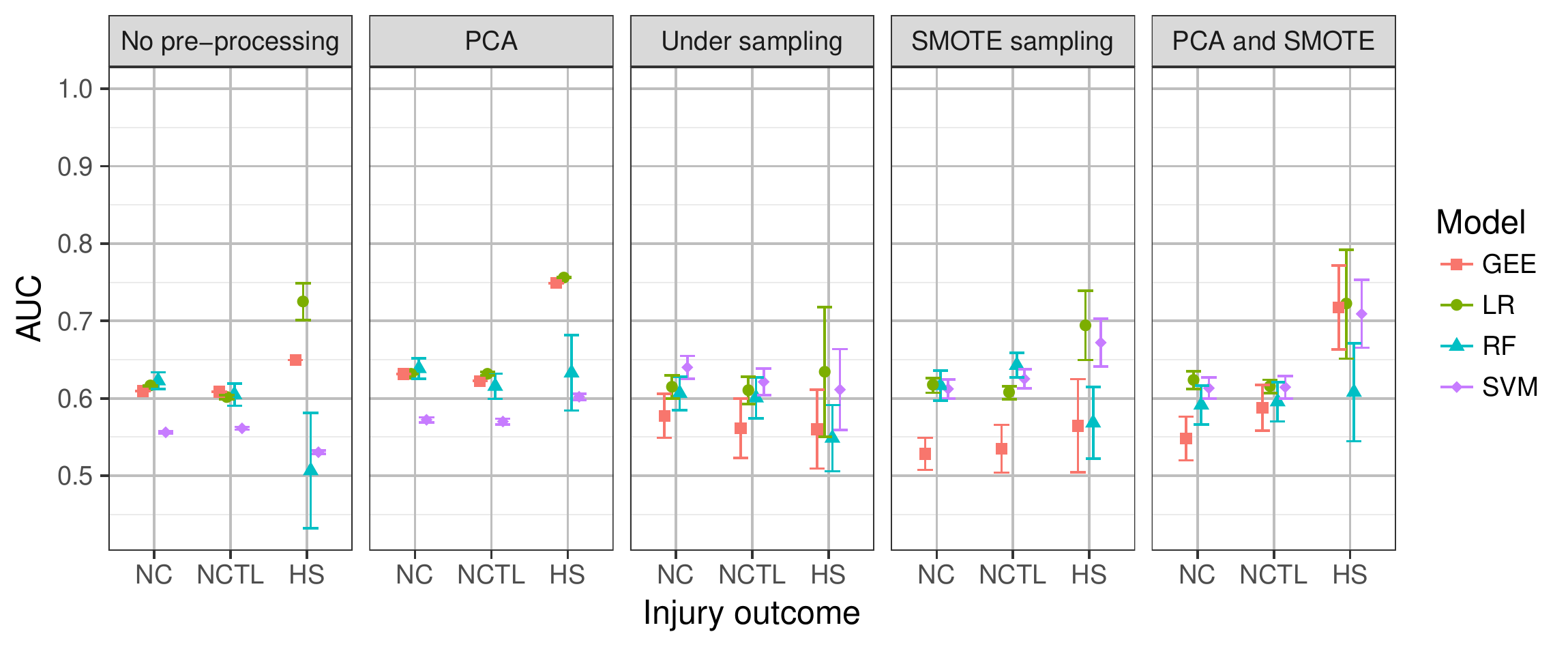}
	\caption{Effects of data pre-processing protocols and sampling methods on multivariate model performance (mean and standard deviation of 50 simulations) for each injury outcome.}
	\label{fig3}
\end{figure}

\subsection{Model performance for new versus returning players}

Predictive accuracy of NC and NCTL injury models did not significantly change when new players and returning players were considered separately (Figure 4). Hamstring models appeared to perform better on returning players, suggesting they may have started to identify patterns leading to hamstring injury in the existing playing group. However, of the 13 total hamstring injuries in the testing data; 10 were to new players and only 3 to returning players. The inflated results for the returning players may be a consequence of the small sample size and not truly representative of expected future performance.

\begin{figure}[h]
	\centering
	\includegraphics[width = 0.8\textwidth]{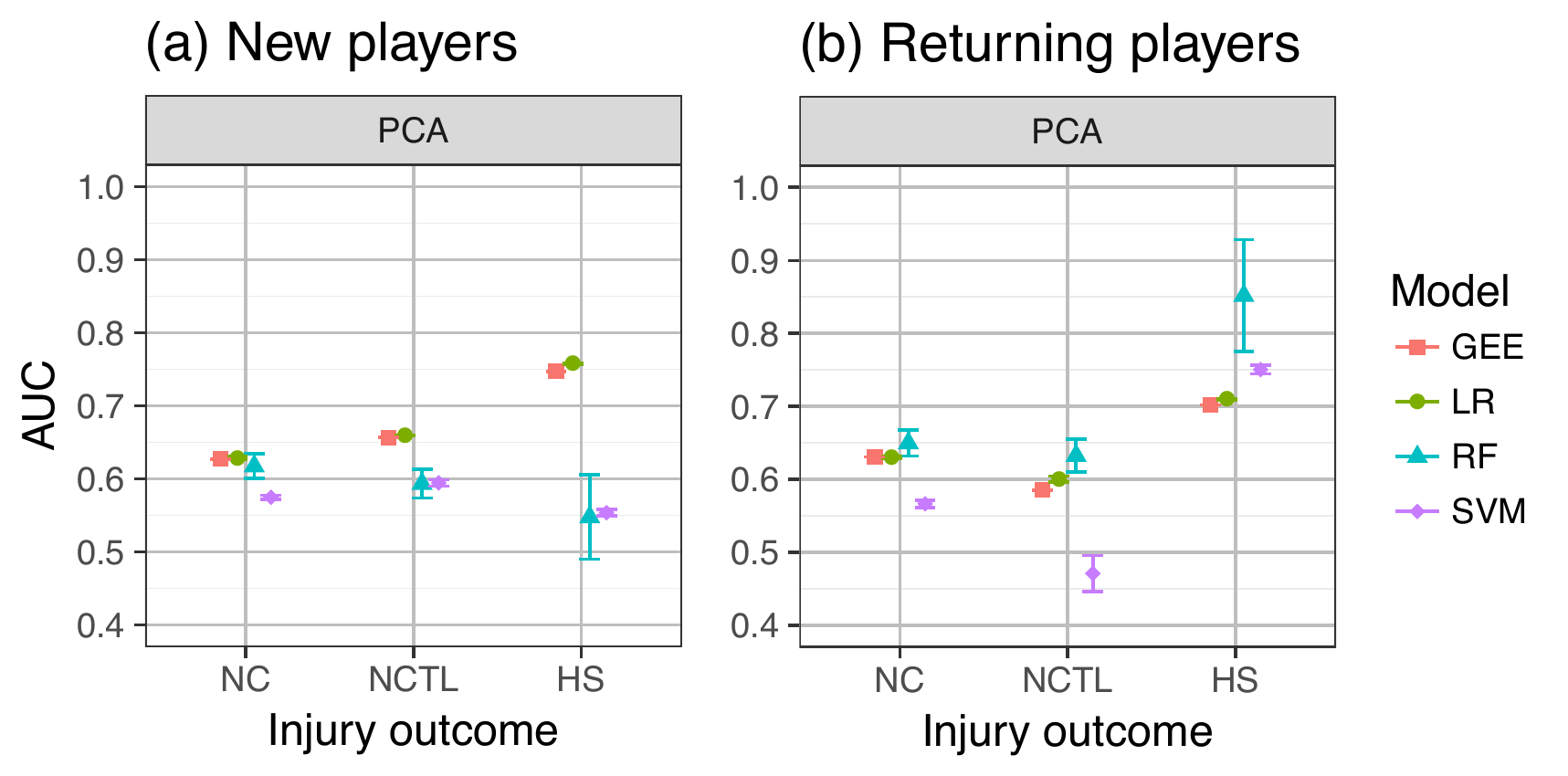}
	\caption{Model performance (mean and standard deviation of 50 simulations) for (a) new players and (b) returning players (PCA pre-processing).}
	\label{fig4}
\end{figure}

\subsection{Effect of increasing the number of model training samples}

The performance of predictive models can be influenced by the amount of data available to build models \cite{kuhn2013}. A learning curve shows the performance of a model on the training and testing data sets as the size of the training data set is increased. This indicates whether performance is improving or plateauing as more data is used. Learning curves were constructed for two injury models (Figure 5) to investigate whether the poor performance could be attributed to insufficient amounts of data.

\begin{figure}[h]
	\centering
	\includegraphics[width = \textwidth]{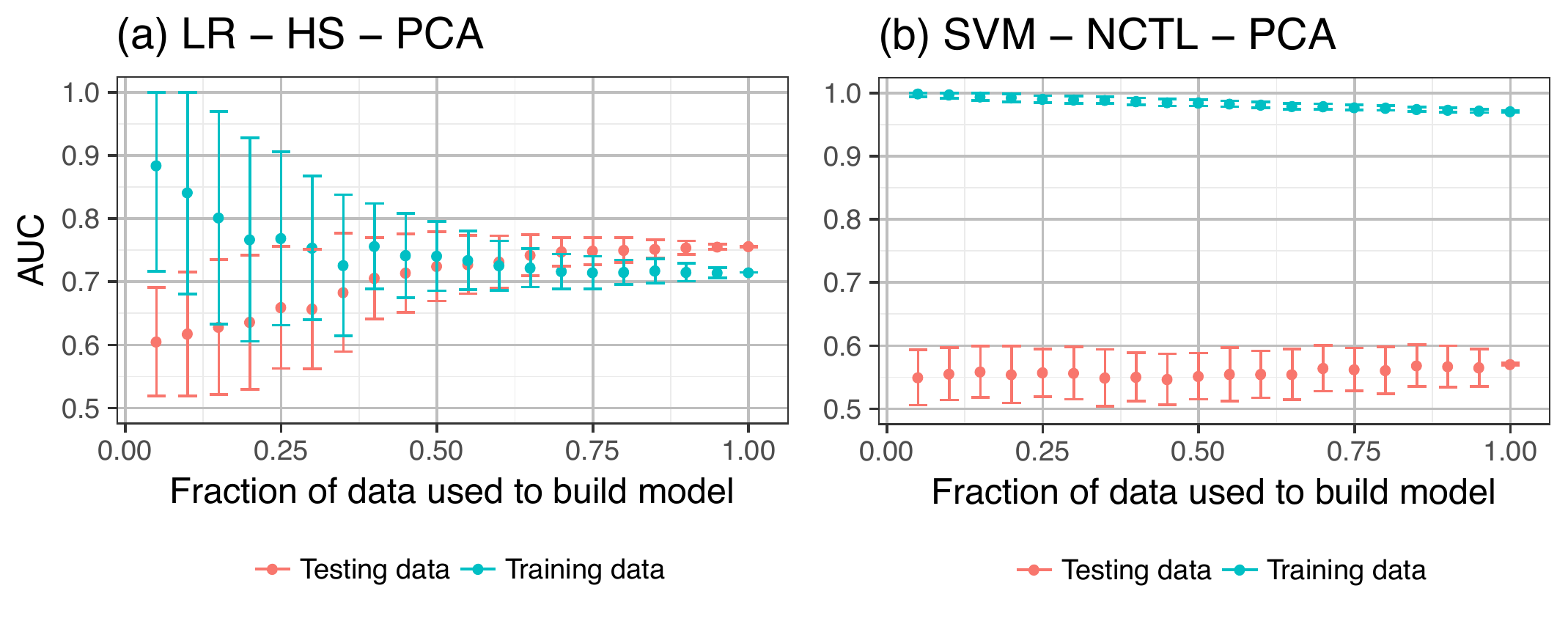}
	\caption{Learning curves showing mean and standard deviation of performance for; (a) LR model of HS injuries (b) SVM model of NCTL injuries (both with PCA pre-processing).}
	\label{fig5}
\end{figure}

The performance of LR to predict hamstring injury (with PCA) is shown in Figure 5(a). Test set performance increased as the amount of data used the build the model increased, suggesting the model was learning to recognise hamstrings injuries better. The performance on the training and testing sets appeared to converge to a similar level (mean AUC=0.7-0.8) once the full data set was used. However this represented a limited ability to predict injuries (Table 3); meaning the model was unable to fully explain the relationship between the predictor and outcome variables, or had underfit the problem. Underfitting suggested the model was unable to capture enough complexity in the data or that the predictor variables did not contain enough information to predict injuries \cite{kuhn2013}.

Figure 5(b) shows the learning curve for an SVM model for NCTL injuries (with PCA). The performance on the model training data was near perfect but test set performance was well below desired levels. The learning curve suggests the model may be suffering from overfitting; it has perfectly fit the injuries in the model training data but does not generalise well to new data. Potential strategies for addressing this are collecting more data (especially more positive injury samples) or penalising the model for increasing complexity (regularisation) \cite{kuhn2013}. The increase in performance observed when the size of the training data set was increased from 460 to 9203 samples was approximately 5\%. Estimating further improvements is difficult given the potentially non-linear relationship, however it appears at least an order of magnitude (tenfold) more data may be needed.

\section{Discussion}

Models of the relationship between training loads and injury in elite Australian footballers showed limited ability to predict future injury. Multivariate models of non-contact and non-contact time-loss injuries performed better than univariate models, but only marginally better than would be expected by random chance (mean AUC$<$0.7). The levels of performance were comparable to similar modeling studies in rugby league (AUC=0.64-0.74) \cite{thornton2016}.

Considering hamstring injuries on their own led to improvements in model performance (best AUC=0.76) (Figure 2-3). Implementing such a model in practice would require practitioners to consider how much modification of player training they are willing to accept in an attempt to prevent injuries. Results suggested that predicting half of the HS injuries would incur a false positive rate above 10\% (or more than 1 in 10 player sessions unnecessarily modified) (Table 3). The multivariate models used in this study provided improved predictive ability compared to findings by Ruddy et. al. \cite{ruddy2016} in a larger cohort of hamstring injuries in Australian football (AUC=0.5-0.63).

Pre-processing data to account for multicollinearity of predictors (with PCA) and sampling to reduce the level of class imbalance resulted in minor improvements to predictive performance (Figure 3). Particularly in the more complex models (SVM) that tended to over-fit the model training data. Consideration of these issues may improve predictive performance in future modelling studies.

\subsection{Possibilities for improving injury prediction models}

The learning curves for different modelling approaches are shown in Figure 5. These provide some indicators for ways to potentially improve the performance of future injury prediction models. Underfitting models (Figure 5(a)) can be improved by increasing the model complexity or by increasing the number and variety of predictors \cite{kuhn2013}. This suggests that linear models such as logistic regression may not be well suited to modeling complex injury relationships, supporting the contention of Bittencourt et. al. \cite{bittencourt2016}. The more complex models (RF and SVM) tended to over-fit the small number of injuries in the data (Figure 5(b)) and did not generalise well to future injury observations. Collecting more player monitoring and injury data ($>$10 seasons) may provide a way to construct practically useful prediction models.

\subsection{Utility of training load monitoring for injury risk reduction}

Results of this study suggested that training loads models were limited in their ability predict future injuries for an athlete on a given day. However this not does not rule out training load monitoring as a valid practice. There is strong evidence \cite{blanch2016,carey2016,hulin2015,malone2016,murray2016,soligard2016} that rapid increases in load and spikes in acute:chronic workload ratio are associated with increases in team injury rates. For this reason, measuring absolute and relative training loads in team sports to monitor load progression and allow for informed modification of plans is still considered best practice \cite{drew2016,soligard2016}. Using individual training loads as a daily decision tool for athlete injury prediction had limited utility in this study.

\subsection{Limitations}

The professional sporting team participating in this study had a high amount of player turnover. This may have impacted on the accuracy of injury prediction models (Figure 4) and restricted the ability to build player specific models. The small injury sample size was another consequence of conducting this study within a single club. There were an insufficient number of injury records of a particular type (other than hamstrings) to create different injury specific models.

This study included a number of the commonly used training load measures (GPS and session-RPE) and derived metrics (cumulative load, acute:chronic workload ratio, monotony and strain). Running demands in relative speed zones, subjective wellness and fatigue markers, and biomechanical screening data were not available. These variables may contain relevant information to improve predictive models.

In both the model training and testing seasons physical preparation staff were aware of the emerging body of research on training loads and injury risk \cite{colby2014,gabbett2010,gabbett2016,hulin2015,rogalski2013} and gave consideration to this when planning training. This may have influenced the distribution of training load variables recorded in this study.

\section{Conclusion}

Models of training load, age and session type showed limited ability to predict future injury in Australian footballers. Hamstring injury specific models showed potential for better performance than general non-contact injury models. Performance tended to improve with increasing quantity of data, highlighting the limitations of predictive modelling attempts conducted within a single team. Training load may be an import injury risk factor to monitor, but is limited as a daily decision tool for injury prediction within a single Australian football club.

\newpage
\section*{Appendix I: Methods of training load calculation}
\begin{itemize}

\item Rolling averages ($C$) of workload ($w$) using a 3, 6 and 21 day accumulation window were calculated on each day ($i$) of the season. These time periods have been identified as appropriate for quantifying cumulative load in Australian football \cite{colby2014,carey2016}.

\begin{equation*}
C_i = \sum_{j = i-c}^{i-1} \frac{w_j}{c} \quad \text{for} \quad  c \in {3, 6, 21}
\end{equation*}

\item Exponentially weighted moving averages ($EWMA$) were calculated daily with 3, 6 and 21 day decay parameters ($N$). An exponentially weighted moving average weights the influence of training loads less the longer ago they happened. The method used was adapted Williams et al. \cite{williams2016better} so that the value calculated on day ($i$) had no dependence on the training load that day ($w_i$). This is necessary to avoid information recorded on the day of an injury being used to try and predict that injury.

\begin{equation*}
EWMA_i = \lambda \cdot w_{i-1} + (1 - \lambda) \cdot EWMA_{i-1}
\end{equation*}
\begin{equation*}
\lambda = \frac{2}{N + 1} \quad \text{for} \quad N \in {3,6,21}
\end{equation*}

\item Training monotony was calculated each day as the average training load in the previous 7 days divided by the standard deviation of daily loads over the same time \cite{foster2001}. Monotony represents the variation in training done by an athlete, with higher values indicating more monotonous training. Monotony was not calculated for HSR due to players frequently accumulating zero HSR load in the previous 7 days.

\item Training strain was calculated as the sum of load in the previous 7 days multiplied by the training monotony. Strain is an extension of cumulative training volume that incorporates a weighting factor based on the amount of daily variation \cite{foster2001}. Strain was not calculated for HSR due to players frequently accumulating zero HSR load in the previous 7 days.

\item Daily acute:chronic workload ratios ($r$) were derived for each  workload variable; 3 and 6 day acute time windows ($a$) were used with a 21 day chronic window ($c$). These choices have been identified as appropriate for discriminating between high and low risk athletes in the study cohort \cite{carey2016}. When players had no chronic workload (and by definition zero acute load) they were assigned an acute:chronic workload ratio of zero.

\begin{equation*}
r_i = \sum_{j = i - a}^{i-1} \frac{w_j}{a} / \sum_{j = i - c}^{i-1} \frac{w_j}{c}
\end{equation*}

\item Exponentially weighted acute:chronic workload ratios were included as a modification of the acute:chronic workload ratio where the rolling averages were replaced by exponentially weighted moving averages \cite{williams2016better,murray2016}. Murray et al. suggested that the exponentially weighted acute:chronic workload ratio gave a better indicator of injury risk than the rolling average method \cite{murray2016}. 3 and 6 day acute and 21 day chronic decay parameters were used.

\end{itemize}

\newpage

\section*{Supplementary table I: Descriptive statistics for model training and testing data}

\begin{longtable}{|c|ccc|ccc|c|}
	\hline
	&                               & \begin{tabular}[c]{@{}c@{}}Testing\\ data\end{tabular}                        &                                                          &                              & \begin{tabular}[c]{@{}c@{}}Model\\ training\\ data\end{tabular}               &                                                          &                                                         \\ \hline
	Variable                                                             & \multicolumn{1}{c|}{Median}   & \multicolumn{1}{c|}{\begin{tabular}[c]{@{}c@{}}First\\ Quartile\end{tabular}} & \begin{tabular}[c]{@{}c@{}}Third\\ Quartile\end{tabular} & \multicolumn{1}{c|}{Median}  & \multicolumn{1}{c|}{\begin{tabular}[c]{@{}c@{}}First\\ Quartile\end{tabular}} & \begin{tabular}[c]{@{}c@{}}Third\\ Quartile\end{tabular} & \begin{tabular}[c]{@{}c@{}}Effect\\ Size$^{\#}$\end{tabular} \\ \hline
	EWMA Distance 3                                                      & \multicolumn{1}{c|}{1819.61}  & \multicolumn{1}{c|}{1445.46}                                                  & 2171.71                                                  & \multicolumn{1}{c|}{1822.17} & \multicolumn{1}{c|}{1378.66}                                                  & 2292.29                                                  & 0.01                                                    \\ \hline
	EWMA MSR 3                                                           & \multicolumn{1}{c|}{187.82}   & \multicolumn{1}{c|}{134.53}                                                   & 256.99                                                   & \multicolumn{1}{c|}{189.76}  & \multicolumn{1}{c|}{124.04}                                                   & 280.47                                                   & 0.01                                                    \\ \hline
	EWMA HSR 3                                                           & \multicolumn{1}{c|}{31.41}    & \multicolumn{1}{c|}{17.44}                                                    & 49.05                                                    & \multicolumn{1}{c|}{24.29}   & \multicolumn{1}{c|}{12.1}                                                     & 42                                                       & -0.16                                                   \\ \hline
	\begin{tabular}[c]{@{}c@{}}EWMA\\ Player Load 3\end{tabular}         & \multicolumn{1}{c|}{184.36}   & \multicolumn{1}{c|}{145.84}                                                   & 225.1                                                    & \multicolumn{1}{c|}{192.45}  & \multicolumn{1}{c|}{142.99}                                                   & 246.85                                                   & 0.06                                                    \\ \hline
	EWMA sRPE 3                                                          & \multicolumn{1}{c|}{106.29}   & \multicolumn{1}{c|}{77.22}                                                    & 141.26                                                   & \multicolumn{1}{c|}{107.79}  & \multicolumn{1}{c|}{75.38}                                                    & 147.38                                                   & 0                                                       \\ \hline
	EWMA Distance 6                                                      & \multicolumn{1}{c|}{2321.25}  & \multicolumn{1}{c|}{1946.34}                                                  & 2643.48                                                  & \multicolumn{1}{c|}{2342.01} & \multicolumn{1}{c|}{1910.63}                                                  & 2788.72                                                  & 0.03                                                    \\ \hline
	EWMA MSR 6                                                           & \multicolumn{1}{c|}{254.46}   & \multicolumn{1}{c|}{202.88}                                                   & 331.09                                                   & \multicolumn{1}{c|}{259.82}  & \multicolumn{1}{c|}{193.72}                                                   & 349.42                                                   & 0.01                                                    \\ \hline
	EWMA HSR 6                                                           & \multicolumn{1}{c|}{44.74}    & \multicolumn{1}{c|}{30.32}                                                    & 61.9                                                     & \multicolumn{1}{c|}{35.76}   & \multicolumn{1}{c|}{22.58}                                                    & 53.62                                                    & -0.19                                                   \\ \hline
	\begin{tabular}[c]{@{}c@{}}EWMA\\ Player Load 6\end{tabular}         & \multicolumn{1}{c|}{234.03}   & \multicolumn{1}{c|}{195.96}                                                   & 273.19                                                   & \multicolumn{1}{c|}{247.54}  & \multicolumn{1}{c|}{199.32}                                                   & 300.16                                                   & 0.11                                                    \\ \hline
	EWMA sRPE 6                                                          & \multicolumn{1}{c|}{146.28}   & \multicolumn{1}{c|}{119.25}                                                   & 180.19                                                   & \multicolumn{1}{c|}{146.44}  & \multicolumn{1}{c|}{113.92}                                                   & 187.7                                                    & 0                                                       \\ \hline
	EWMA Distance 21                                                     & \multicolumn{1}{c|}{2741.26}  & \multicolumn{1}{c|}{2454.06}                                                  & 3002.16                                                  & \multicolumn{1}{c|}{2719.93} & \multicolumn{1}{c|}{2400.07}                                                  & 3037.63                                                  & -0.01                                                   \\ \hline
	EWMA MSR:21                                                          & \multicolumn{1}{c|}{317.65}   & \multicolumn{1}{c|}{274.2}                                                    & 373.27                                                   & \multicolumn{1}{c|}{313.99}  & \multicolumn{1}{c|}{257.69}                                                   & 379.16                                                   & -0.04                                                   \\ \hline
	EWMA HSR 21                                                          & \multicolumn{1}{c|}{55.22}    & \multicolumn{1}{c|}{42.41}                                                    & 69.25                                                    & \multicolumn{1}{c|}{44.42}   & \multicolumn{1}{c|}{31.7}                                                     & 58.25                                                    & -0.28                                                   \\ \hline
	\begin{tabular}[c]{@{}c@{}}EWMA\\ Player Load 21\end{tabular}        & \multicolumn{1}{c|}{275.59}   & \multicolumn{1}{c|}{241.64}                                                   & 305.72                                                   & \multicolumn{1}{c|}{284.81}  & \multicolumn{1}{c|}{247.6}                                                    & 324.8                                                    & 0.13                                                    \\ \hline
	EWMA sRPE 21                                                         & \multicolumn{1}{c|}{181.11}   & \multicolumn{1}{c|}{160.17}                                                   & 200.53                                                   & \multicolumn{1}{c|}{175.71}  & \multicolumn{1}{c|}{151.76}                                                   & 200.48                                                   & -0.07                                                   \\ \hline
	\begin{tabular}[c]{@{}c@{}}EWMA ACWR\\ Distance 3:21\end{tabular}    & \multicolumn{1}{c|}{0.67}     & \multicolumn{1}{c|}{0.55}                                                     & 0.81                                                     & \multicolumn{1}{c|}{0.67}    & \multicolumn{1}{c|}{0.54}                                                     & 0.84                                                     & 0                                                       \\ \hline
	\begin{tabular}[c]{@{}c@{}}EWMA ACWR\\ MSR 3:21\end{tabular}         & \multicolumn{1}{c|}{0.6}      & \multicolumn{1}{c|}{0.43}                                                     & 0.77                                                     & \multicolumn{1}{c|}{0.61}    & \multicolumn{1}{c|}{0.43}                                                     & 0.84                                                     & 0.04                                                    \\ \hline
	\begin{tabular}[c]{@{}c@{}}EWMA ACWR\\ HSR 3:21\end{tabular}         & \multicolumn{1}{c|}{0.58}     & \multicolumn{1}{c|}{0.36}                                                     & 0.84                                                     & \multicolumn{1}{c|}{0.57}    & \multicolumn{1}{c|}{0.33}                                                     & 0.87                                                     & -0.02                                                   \\ \hline
	\begin{tabular}[c]{@{}c@{}}EWMA ACWR\\ Player Load 3:21\end{tabular} & \multicolumn{1}{c|}{0.68}     & \multicolumn{1}{c|}{0.56}                                                     & 0.83                                                     & \multicolumn{1}{c|}{0.67}    & \multicolumn{1}{c|}{0.54}                                                     & 0.84                                                     & -0.01                                                   \\ \hline
	\begin{tabular}[c]{@{}c@{}}EWMA ACWR\\ sRPE 3:21\end{tabular}        & \multicolumn{1}{c|}{0.61}     & \multicolumn{1}{c|}{0.46}                                                     & 0.78                                                     & \multicolumn{1}{c|}{0.62}    & \multicolumn{1}{c|}{0.47}                                                     & 0.81                                                     & 0.03                                                    \\ \hline
	\begin{tabular}[c]{@{}c@{}}EWMA ACWR\\ Distance 6:21\end{tabular}    & \multicolumn{1}{c|}{0.85}     & \multicolumn{1}{c|}{0.77}                                                     & 0.94                                                     & \multicolumn{1}{c|}{0.87}    & \multicolumn{1}{c|}{0.77}                                                     & 0.98                                                     & 0.05                                                    \\ \hline
	\begin{tabular}[c]{@{}c@{}}EWMA ACWR\\ MSR 6:21\end{tabular}         & \multicolumn{1}{c|}{0.81}     & \multicolumn{1}{c|}{0.69}                                                     & 0.94                                                     & \multicolumn{1}{c|}{0.83}    & \multicolumn{1}{c|}{0.7}                                                      & 0.99                                                     & 0.06                                                    \\ \hline
	\begin{tabular}[c]{@{}c@{}}EWMA ACWR\\ HSR 6:21\end{tabular}         & \multicolumn{1}{c|}{0.82}     & \multicolumn{1}{c|}{0.66}                                                     & 1                                                        & \multicolumn{1}{c|}{0.82}    & \multicolumn{1}{c|}{0.64}                                                     & 1.04                                                     & 0.02                                                    \\ \hline
	\begin{tabular}[c]{@{}c@{}}EWMA ACWR\\ Player Load 6:21\end{tabular} & \multicolumn{1}{c|}{0.86}     & \multicolumn{1}{c|}{0.78}                                                     & 0.95                                                     & \multicolumn{1}{c|}{0.86}    & \multicolumn{1}{c|}{0.77}                                                     & 0.98                                                     & 0.04                                                    \\ \hline
	\begin{tabular}[c]{@{}c@{}}EWMA ACWR\\ sRPE 6:21\end{tabular}        & \multicolumn{1}{c|}{0.82}     & \multicolumn{1}{c|}{0.71}                                                     & 0.95                                                     & \multicolumn{1}{c|}{0.84}    & \multicolumn{1}{c|}{0.72}                                                     & 0.99                                                     & 0.05                                                    \\ \hline
	Monotony Distance                                                    & \multicolumn{1}{c|}{0.68}     & \multicolumn{1}{c|}{0.6}                                                      & 0.75                                                     & \multicolumn{1}{c|}{0.69}    & \multicolumn{1}{c|}{0.58}                                                     & 0.76                                                     & 0.02                                                    \\ \hline
	Monotony MSR                                                         & \multicolumn{1}{c|}{0.58}     & \multicolumn{1}{c|}{0.53}                                                     & 0.64                                                     & \multicolumn{1}{c|}{0.6}     & \multicolumn{1}{c|}{0.54}                                                     & 0.7                                                      & 0.15                                                    \\ \hline
	\begin{tabular}[c]{@{}c@{}}Monotony\\ Player Load\end{tabular}       & \multicolumn{1}{c|}{0.69}     & \multicolumn{1}{c|}{0.61}                                                     & 0.76                                                     & \multicolumn{1}{c|}{0.69}    & \multicolumn{1}{c|}{0.58}                                                     & 0.76                                                     & -0.03                                                   \\ \hline
	Monotony sRPE                                                        & \multicolumn{1}{c|}{0.59}     & \multicolumn{1}{c|}{0.54}                                                     & 0.66                                                     & \multicolumn{1}{c|}{0.6}     & \multicolumn{1}{c|}{0.54}                                                     & 0.71                                                     & 0.07                                                    \\ \hline
	Strain Distance                                                      & \multicolumn{1}{c|}{12989.53} & \multicolumn{1}{c|}{10114.39}                                                 & 15957.49                                                 & \multicolumn{1}{c|}{13459.2} & \multicolumn{1}{c|}{9872.34}                                                  & 17586.5                                                  & 0.07                                                    \\ \hline
	Strain MSR                                                           & \multicolumn{1}{c|}{1274.41}  & \multicolumn{1}{c|}{940.68}                                                   & 1619                                                     & \multicolumn{1}{c|}{1361.02} & \multicolumn{1}{c|}{941}                                                      & 1916.51                                                  & 0.09                                                    \\ \hline
	Strain Player Load                                                   & \multicolumn{1}{c|}{1325.09}  & \multicolumn{1}{c|}{1014.15}                                                  & 1649.67                                                  & \multicolumn{1}{c|}{1424.26} & \multicolumn{1}{c|}{1030}                                                     & 1866.05                                                  & 0.12                                                    \\ \hline
	Strain sRPE                                                          & \multicolumn{1}{c|}{762.03}   & \multicolumn{1}{c|}{574.8}                                                    & 921.85                                                   & \multicolumn{1}{c|}{778.46}  & \multicolumn{1}{c|}{571.98}                                                   & 1039.11                                                  & 0.07                                                    \\ \hline
	Mean Distance 3                                                      & \multicolumn{1}{c|}{2195.75}  & \multicolumn{1}{c|}{1472.29}                                                  & 3115.33                                                  & \multicolumn{1}{c|}{2240.33} & \multicolumn{1}{c|}{1447.56}                                                  & 3265.33                                                  & 0.01                                                    \\ \hline
	Mean MSR 3                                                           & \multicolumn{1}{c|}{240.78}   & \multicolumn{1}{c|}{128.08}                                                   & 400                                                      & \multicolumn{1}{c|}{244}     & \multicolumn{1}{c|}{125}                                                      & 412.33                                                   & 0.01                                                    \\ \hline
	Mean HSR 3                                                           & \multicolumn{1}{c|}{43}       & \multicolumn{1}{c|}{11.29}                                                    & 74.91                                                    & \multicolumn{1}{c|}{30.33}   & \multicolumn{1}{c|}{4.34}                                                     & 61                                                       & -0.13                                                   \\ \hline
	Mean Player Load 3                                                   & \multicolumn{1}{c|}{226.04}   & \multicolumn{1}{c|}{150.25}                                                   & 315.25                                                   & \multicolumn{1}{c|}{235.67}  & \multicolumn{1}{c|}{148.48}                                                   & 351                                                      & 0.04                                                    \\ \hline
	Mean sRPE 3                                                          & \multicolumn{1}{c|}{134.67}   & \multicolumn{1}{c|}{76.46}                                                    & 207.57                                                   & \multicolumn{1}{c|}{134.75}  & \multicolumn{1}{c|}{71.67}                                                    & 225.94                                                   & 0                                                       \\ \hline
	Mean Distance 6                                                      & \multicolumn{1}{c|}{2695.72}  & \multicolumn{1}{c|}{1996}                                                     & 3250.83                                                  & \multicolumn{1}{c|}{2683.17} & \multicolumn{1}{c|}{2003.05}                                                  & 3233.29                                                  & 0                                                       \\ \hline
	Mean MSR 6                                                           & \multicolumn{1}{c|}{307.01}   & \multicolumn{1}{c|}{214.01}                                                   & 394.85                                                   & \multicolumn{1}{c|}{299.24}  & \multicolumn{1}{c|}{206}                                                      & 398.33                                                   & -0.01                                                   \\ \hline
	Mean HSR 6                                                           & \multicolumn{1}{c|}{50.03}    & \multicolumn{1}{c|}{31.28}                                                    & 72.93                                                    & \multicolumn{1}{c|}{40.05}   & \multicolumn{1}{c|}{23}                                                       & 62.67                                                    & -0.18                                                   \\ \hline
	Mean Player Load 6                                                   & \multicolumn{1}{c|}{270.15}   & \multicolumn{1}{c|}{201.85}                                                   & 324.33                                                   & \multicolumn{1}{c|}{280.33}  & \multicolumn{1}{c|}{210.1}                                                    & 342.22                                                   & 0.09                                                    \\ \hline
	Mean sRPE 6                                                          & \multicolumn{1}{c|}{181}      & \multicolumn{1}{c|}{121.66}                                                   & 222.18                                                   & \multicolumn{1}{c|}{176.83}  & \multicolumn{1}{c|}{119.86}                                                   & 219.5                                                    & -0.02                                                   \\ \hline
	Mean Distance 21                                                     & \multicolumn{1}{c|}{2846.02}  & \multicolumn{1}{c|}{2531.66}                                                  & 3121.51                                                  & \multicolumn{1}{c|}{2847.81} & \multicolumn{1}{c|}{2474.67}                                                  & 3174.09                                                  & 0.01                                                    \\ \hline
	Mean MSR:21                                                          & \multicolumn{1}{c|}{330.65}   & \multicolumn{1}{c|}{283.76}                                                   & 383.48                                                   & \multicolumn{1}{c|}{325.38}  & \multicolumn{1}{c|}{262.35}                                                   & 397.98                                                   & -0.04                                                   \\ \hline
	Mean HSR 21                                                          & \multicolumn{1}{c|}{57.44}    & \multicolumn{1}{c|}{42.97}                                                    & 72.43                                                    & \multicolumn{1}{c|}{45.88}   & \multicolumn{1}{c|}{31.34}                                                    & 61.11                                                    & -0.27                                                   \\ \hline
	Mean Player Load 21                                                  & \multicolumn{1}{c|}{285.79}   & \multicolumn{1}{c|}{248.85}                                                   & 317.34                                                   & \multicolumn{1}{c|}{297.73}  & \multicolumn{1}{c|}{254}                                                      & 338.78                                                   & 0.14                                                    \\ \hline
	Mean sRPE 21                                                         & \multicolumn{1}{c|}{190.68}   & \multicolumn{1}{c|}{167.44}                                                   & 210.15                                                   & \multicolumn{1}{c|}{184.64}  & \multicolumn{1}{c|}{155.76}                                                   & 210.96                                                   & -0.06                                                   \\ \hline
	ACWR Distance 3:21                                                   & \multicolumn{1}{c|}{0.81}     & \multicolumn{1}{c|}{0.54}                                                     & 1.22                                                     & \multicolumn{1}{c|}{0.82}    & \multicolumn{1}{c|}{0.53}                                                     & 1.26                                                     & 0.01                                                    \\ \hline
	ACWR MSR 3:21                                                        & \multicolumn{1}{c|}{0.76}     & \multicolumn{1}{c|}{0.37}                                                     & 1.25                                                     & \multicolumn{1}{c|}{0.78}    & \multicolumn{1}{c|}{0.39}                                                     & 1.3                                                      & 0.03                                                    \\ \hline
	ACWR HSR 3:21                                                        & \multicolumn{1}{c|}{0.8}      & \multicolumn{1}{c|}{0.24}                                                     & 1.31                                                     & \multicolumn{1}{c|}{0.73}    & \multicolumn{1}{c|}{0.12}                                                     & 1.4                                                      & -0.02                                                   \\ \hline
	\begin{tabular}[c]{@{}c@{}}ACWR\\ Player Load 3:21\end{tabular}      & \multicolumn{1}{c|}{0.84}     & \multicolumn{1}{c|}{0.55}                                                     & 1.21                                                     & \multicolumn{1}{c|}{0.82}    & \multicolumn{1}{c|}{0.53}                                                     & 1.26                                                     & 0                                                       \\ \hline
	ACWR sRPE 3:21                                                       & \multicolumn{1}{c|}{0.73}     & \multicolumn{1}{c|}{0.43}                                                     & 1.24                                                     & \multicolumn{1}{c|}{0.77}    & \multicolumn{1}{c|}{0.42}                                                     & 1.34                                                     & 0.02                                                    \\ \hline
	ACWR Distance 6:21                                                   & \multicolumn{1}{c|}{0.97}     & \multicolumn{1}{c|}{0.76}                                                     & 1.12                                                     & \multicolumn{1}{c|}{0.97}    & \multicolumn{1}{c|}{0.76}                                                     & 1.13                                                     & 0.01                                                    \\ \hline
	ACWR MSR 6:21                                                        & \multicolumn{1}{c|}{0.93}     & \multicolumn{1}{c|}{0.67}                                                     & 1.15                                                     & \multicolumn{1}{c|}{0.94}    & \multicolumn{1}{c|}{0.68}                                                     & 1.16                                                     & 0.02                                                    \\ \hline
	ACWR HSR 6:21                                                        & \multicolumn{1}{c|}{0.91}     & \multicolumn{1}{c|}{0.62}                                                     & 1.23                                                     & \multicolumn{1}{c|}{0.93}    & \multicolumn{1}{c|}{0.6}                                                      & 1.29                                                     & 0.02                                                    \\ \hline
	\begin{tabular}[c]{@{}c@{}}ACWR\\ Player Load 6:21\end{tabular}      & \multicolumn{1}{c|}{0.97}     & \multicolumn{1}{c|}{0.76}                                                     & 1.11                                                     & \multicolumn{1}{c|}{0.96}    & \multicolumn{1}{c|}{0.75}                                                     & 1.13                                                     & 0.01                                                    \\ \hline
	ACWR sRPE 6:21                                                       & \multicolumn{1}{c|}{0.98}     & \multicolumn{1}{c|}{0.7}                                                      & 1.15                                                     & \multicolumn{1}{c|}{0.97}    & \multicolumn{1}{c|}{0.71}                                                     & 1.17                                                     & 0.01                                                    \\ \hline
	Age                                                                  & \multicolumn{1}{c|}{22}       & \multicolumn{1}{c|}{20}                                                       & 27                                                       & \multicolumn{1}{c|}{23}      & \multicolumn{1}{c|}{21}                                                       & 26                                                       & 0.04                                                    \\ \hline
\end{longtable}

$^{\#}$Effect size calculated as the rank biserial correlation from the Mann-Whitney U test \cite{wendt1972}.

\newpage

\bibliography{injPredictArxiv}

\begin{thebibliography}{10}

\bibitem{rogalski2013}
Brent Rogalski, Brian Dawson, Jarryd Heasman, and Tim~J. Gabbett.
\newblock Training and game loads and injury risk in elite australian
  footballers.
\newblock {\em Journal of Science and Medicine in Sport}, 16:499--503, 2013.

\bibitem{colby2014}
Marcus~J. Colby, Brian Dawson, Jarryd Heasman, Brent Rogalski, and Tim~J.
  Gabbett.
\newblock Accelerometer and gps-derived running loads and injury risk in elite
  australian footballers.
\newblock {\em The Journal of Strength \& Conditioning Research},
  28:2244–2252, 2014.

\bibitem{hulin2015}
Billy~T Hulin, Tim~J Gabbett, Daniel~W Lawson, Peter Caputi, and John~A
  Sampson.
\newblock The acute:chronic workload ratio predicts injury: high chronic
  workload may decrease injury risk in elite rugby league players.
\newblock {\em British Journal of Sports Medicine}, pages
  bjsports--2015--094817, 2015.

\bibitem{carey2016}
David~L Carey, Peter Blanch, Kok-Leong Ong, Kay~M Crossley, Justin Crow, and
  Meg~E Morris.
\newblock Training loads and injury risk in australian football - differing
  acute:chronic workload ratios influence match injury risk.
\newblock {\em British Journal of Sports Medicine}, pages
  bjsports--2016--096309, 2016.

\bibitem{malone2016}
Shane Malone, Adam Owen, Matt Newton, Bruno Mendes, Kieran~D. Collins, and
  Tim~J. Gabbett.
\newblock The acute:chonic workload ratio in relation to injury risk in
  professional soccer.
\newblock {\em Journal of Science and Medicine in Sport}, 2016.

\bibitem{murray2016}
Nicholas~B Murray, Tim~J Gabbett, Andrew~D Townshend, and Peter Blanch.
\newblock Calculating acute:chronic workload ratios using exponentially
  weighted moving averages provides a more sensitive indicator of injury
  likelihood than rolling averages.
\newblock {\em British Journal of Sports Medicine}, pages
  bjsports--2016--097152, 2016.

\bibitem{thornton2016}
Heidi~R Thornton, Jace~A Delaney, Grant~M Duthie, and Ben~J Dascombe.
\newblock Importance of various training load measures on injury incidence of
  professional rugby league athletes.
\newblock {\em International Journal of Sports Physiology and Performance},
  pages 1--17, 2016.

\bibitem{soligard2016}
Torbjørn Soligard, Martin Schwellnus, Juan-Manuel Alonso, Roald Bahr, Ben
  Clarsen, H~Paul Dijkstra, Tim Gabbett, Michael Gleeson, Martin Hägglund,
  Mark~R Hutchinson, Christa Janse~van Rensburg, Karim~M Khan, Romain Meeusen,
  John~W Orchard, Babette~M Pluim, Martin Raftery, Richard Budgett, and Lars
  Engebretsen.
\newblock How much is too much? (part 1) international olympic committee
  consensus statement on load in sport and risk of injury.
\newblock {\em British Journal of Sports Medicine}, 50:1030--1041, 2016.

\bibitem{hagglund2013}
Martin Hägglund, Markus Waldén, Henrik Magnusson, Karolina Kristenson, Håkan
  Bengtsson, and Jan Ekstrand.
\newblock Injuries affect team performance negatively in professional football:
  an 11-year follow-up of the uefa champions league injury study.
\newblock {\em British journal of sports medicine}, 47(12):738--742, 2013.

\bibitem{drew2016}
Michael~K. Drew and Caroline~F. Finch.
\newblock The relationship between training load and injury, illness and
  soreness: A systematic and literature review.
\newblock {\em Sports Medicine}, 2016.

\bibitem{blanch2016}
Peter Blanch and Tim~J Gabbett.
\newblock Has the athlete trained enough to return to play safely? the acute:
  chronic workload ratio permits clinicians to quantify a player's risk of
  subsequent injury.
\newblock {\em British journal of sports medicine}, 50(8):471--475, 2016.

\bibitem{gabbett2016}
Tim~J Gabbett.
\newblock The training—injury prevention paradox: should athletes be training
  smarter and harder?
\newblock {\em British Journal of Sports Medicine}, 50:273--280, 2016.

\bibitem{williams2016better}
Sean Williams, Stephen West, Matthew~J Cross, and Keith~A Stokes.
\newblock Better way to determine the acute:chronic workload ratio?
\newblock {\em British Journal of Sports Medicine}, pages
  bjsports--2016--096589, 2016.

\bibitem{foster2001}
Carl Foster, Jessica~A. Florhaug, Jodi Franklin, Lori Gottschall, Lauri~A.
  Hrovatin, Suzanne Parker, Pamela Doleshal, and Christopher Dodge.
\newblock A new approach to monitoring exercise training.
\newblock {\em The Journal of Strength \& Conditioning Research}, 15:109–115,
  2001.

\bibitem{bittencourt2016}
NFN Bittencourt, WH~Meeuwisse, LD~Mendonça, A~Nettel-Aguirre, JM~Ocarino, and
  ST~Fonseca.
\newblock Complex systems approach for sports injuries: moving from risk factor
  identification to injury pattern recognition—narrative review and new
  concept.
\newblock {\em British journal of sports medicine}, 50(21):1309--1314, 2016.

\bibitem{gabbett2010}
Tim~J. Gabbett.
\newblock The development and application of an injury prediction model for
  noncontact, soft-tissue injuries in elite collision sport athletes.
\newblock {\em The Journal of Strength \& Conditioning Research},
  24:2593–2603, 2010.

\bibitem{boyd2011}
Luke~J Boyd, Kevin Ball, and Robert~J Aughey.
\newblock The reliability of minimaxx accelerometers for measuring physical
  activity in australian football.
\newblock {\em International Journal of Sports Physiology and Performance},
  6(3):311--321, 2011.

\bibitem{rampinini2015}
E.~Rampinini, G.~Alberti, M.~Fiorenza, M.~Riggio, R.~Sassi, T.~O. Borges, and
  A.~J. Coutts.
\newblock Accuracy of gps devices for measuring high-intensity running in
  field-based team sports.
\newblock {\em International Journal of Sports Medicine}, 36(1):49--53, 2015.

\bibitem{buuren2011}
Stef Buuren and Karin Groothuis-Oudshoorn.
\newblock mice: Multivariate imputation by chained equations in r.
\newblock {\em Journal of statistical software}, 45(3), 2011.

\bibitem{rae2007}
Katherine Rae and John Orchard.
\newblock The orchard sports injury classification system (osics) version 10.
\newblock {\em Clinical Journal of Sport Medicine}, 17(3):201--204, 2007.

\bibitem{kuhn2013}
Max Kuhn and Kjell Johnson.
\newblock {\em Applied Predictive Modeling}.
\newblock Springer New York, New York, NY, 2013.

\bibitem{orchard2013}
John~W Orchard, Hugh Seward, and Jessica~J Orchard.
\newblock Results of 2 decades of injury surveillance and public release of
  data in the australian football league.
\newblock {\em The American journal of sports medicine}, page 0363546513476270,
  2013.

\bibitem{zou2005}
Hui Zou and Trevor Hastie.
\newblock Regularization and variable selection via the elastic net.
\newblock {\em Journal of the Royal Statistical Society: Series B (Statistical
  Methodology)}, 67(2):301--320, 2005.

\bibitem{breiman2001}
Leo Breiman.
\newblock Random forests.
\newblock {\em Machine learning}, 45(1):5--32, 2001.

\bibitem{liang1986}
Kung-Yee Liang and Scott~L Zeger.
\newblock Longitudinal data analysis using generalized linear models.
\newblock {\em Biometrika}, 73(1):13--22, 1986.

\bibitem{williamson1996}
Denise~Svendsgaard Williamson, Shrikant~I Bangdiwala, Stephen~W Marshall, and
  Anna~E Waller.
\newblock Repeated measures analysis of binary outcomes: applications to injury
  research.
\newblock {\em Accident Analysis \& Prevention}, 28(5):571--579, 1996.

\bibitem{vapnik1998}
Vladimir~Naumovich Vapnik and Vlamimir Vapnik.
\newblock {\em Statistical learning theory}, volume~1.
\newblock Wiley New York, 1998.

\bibitem{rcore}
R~Core Team.
\newblock R language definition.
\newblock {\em Vienna, Austria: R foundation for statistical computing}, 2000.

\bibitem{williams2016}
Sean Williams, Grant Trewartha, Matthew~J Cross, Simon~PT Kemp, and Keith~A
  Stokes.
\newblock Monitoring what matters: A systematic process for selecting training
  load measures.
\newblock {\em International Journal of Sports Physiology and Performance},
  pages 1--20, 2016.

\bibitem{chawla2002}
Nitesh~V. Chawla, Kevin~W. Bowyer, Lawrence~O. Hall, and W.~Philip Kegelmeyer.
\newblock Smote: synthetic minority over-sampling technique.
\newblock {\em Journal of artificial intelligence research}, 16:321--357, 2002.

\bibitem{ruddy2016}
Joshua~D Ruddy, Christopher~W Pollard, Ryan~G Timmins, Morgan~D Williams,
  Anthony~J Shield, and David~A Opar.
\newblock Running exposure is associated with the risk of hamstring strain
  injury in elite australian footballers.
\newblock {\em British Journal of Sports Medicine}, pages
  bjsports--2016--096777, 2016.

\bibitem{wendt1972}
Hans~W Wendt.
\newblock Dealing with a common problem in social science: A simplified
  rank-biserial coefficient of correlation based on the u statistic.
\newblock {\em European Journal of Social Psychology}, 2(4):463--465, 1972.

\end{thebibliography}
\bibliographystyle{unsrt}

\end{document}